\pdfoutput=1

\documentclass[11pt]{article}

\usepackage[preprint]{acl} 
\usepackage{amsmath}
\usepackage{booktabs} 
\usepackage{times}
\usepackage{latexsym}
\usepackage{float}
\usepackage{amsfonts}
\usepackage{multirow}
\usepackage{hyperref}
\usepackage{enumitem}

\usepackage[T1]{fontenc}

\usepackage[utf8]{inputenc}

\usepackage{microtype}

\usepackage{inconsolata}

\usepackage{graphicx}

\title{ Sequence-aware Large Language Models for Explainable Recommendation} 

\author{
  \textbf{Gangyi Zhang}$^{1,}$\thanks{Equal contribution.} \quad 
  \textbf{Runzhe Teng}$^{1,}$\footnotemark[1] \quad 
  \textbf{Chongming Gao}$^{1,}$\thanks{Corresponding author.} \\
  $^{1}$University of Science and Technology of China \\
  \texttt{gangyi.zhang@outlook.com, \{trz177, chongminggao\}@mail.ustc.edu.cn}
}

\begin{document}
\maketitle
\begin{abstract}
Large Language Models (LLMs) have shown strong potential in generating natural language explanations for recommender systems. However, existing methods often overlook the sequential dynamics of user behavior and rely on evaluation metrics misaligned with practical utility. We propose \textbf{SELLER} (\underline{\textbf{SE}}quence-aware \underline{\textbf{LL}}M-based framework for \underline{\textbf{E}}xplainable \underline{\textbf{R}}ecommendation), which integrates explanation generation with utility-aware evaluation. SELLER combines a dual-path encoder—capturing both user behavior and item semantics—with a Mixture-of-Experts adapter to align these signals with LLMs. A unified evaluation framework assesses explanations via both textual quality and their effect on recommendation outcomes. Experiments on public benchmarks show that SELLER consistently outperforms prior methods in explanation quality and real-world utility. The dataset and code are available at \url{https://github.com/gangyizh/SELLER}. 

\end{abstract}

\section{Introduction}

Recommender systems help users navigate overwhelming information in domains like e-commerce and social media. However, their opaque decision-making often undermines user trust. Explainable Recommendation (ER) aims to improve transparency by providing human-interpretable justifications for recommended items~\cite{10.1561/1500000066,Vultureanu-Albisi2021Recommender}.

Recent advances in Large Language Models (LLMs) have enabled more natural and informative explanations~\cite{ma2024xreclargelanguagemodels,10.1007/978-981-97-5569-1_18}. Existing LLM-based ER approaches can be broadly categorized into: (1) \textit{ID-based} methods~\cite{li2021personalized}, which prompt LLMs with user/item identifiers; (2) \textit{Embedding-based} methods~\cite{li2023personalized}, which integrate latent vectors into prompts; and (3) \textit{Hybrid} methods~\cite{ma2024xreclargelanguagemodels}, which combine collaborative signals and structured metadata. Despite their effectiveness, these methods face two key limitations.

First, they neglect the sequential nature of user behavior. Most models rely on static inputs and fail to capture the evolving preferences reflected in user trajectories. For instance, when a user shifts from ``action'' to ``sci-fi thrillers'', current explanations often reflect item properties \cite{ma2024xreclargelanguagemodels} rather than this preference transition, limiting personalization and user alignment.

Second, current evaluation protocols are misaligned with real-world utility. Metrics like BLEU or BERTScore focus on surface-level similarity, but do not assess whether explanations help users make better decisions \cite{he-etal-2023-blind}. Moreover, using review texts as ground truth \cite{10.1145/3397271.3401281} emphasizes item features rather than users' actual decision rationale.

To address these challenges, we propose \textbf{SELLER}, a \underline{\textbf{SE}}quence-aware \underline{\textbf{LL}}M-based framework for \underline{\textbf{E}}xplainable \underline{\textbf{R}}ecommendation. SELLER features (1) a dual-path encoder that models both collaborative and semantic user behavior, enhanced by a Mixture-of-Experts (MoE) adapter for dynamic integration with the LLM; and (2) a unified evaluation framework that complements text-based metrics with utility-based measures derived from downstream recommendation performance.

Experiments on public benchmarks demonstrate that SELLER achieves superior performance in both explanation quality and practical utility. Our main contributions are:
\begin{enumerate}[topsep=0pt,itemsep=0pt,parsep=0pt,leftmargin=*]
    \item We propose a novel dual-path sequence encoder and MoE adapter that effectively model dynamic user preferences for personalized explanation generation.
    \item We propose a unified evaluation framework that jointly considers linguistic quality and recommendation utility.
    \item Empirical results validating SELLER's advantages over prior methods in both explanation generation and decision support effectiveness.
\end{enumerate}

\section{Preliminaries}
\label{sec:preliminaries}

In this section, we formally define the core task addressed by our framework and introduce our unified evaluation methodology. Unlike traditional methods that focus solely on explanation generation, our approach aims to both generate high-quality sequence-aware explanations and provide a robust framework for assessing their practical utility.

\subsection{Notations and Definitions}
Let $\mathcal{U}=\{u_1, u_2, \ldots, u_N\}$ denote the set of users and 
$\mathcal{I}=\{i_1, i_2, \ldots, i_M\}$ the set of items. 
Each user $u\in\mathcal{U}$ is associated with a time-ordered interaction sequence
\begin{equation}
    S_{(u)} = [\,i_1, i_2, \ldots, i_t\,],
\end{equation}
where $i_t$ corresponds to an item that $u$ has interacted with at time step $t$ (e.g., by clicking, purchasing, or rating). 
In addition, every item $i\in\mathcal{I}$ has a textual description $d_i$ (e.g., product metadata or review snippets), providing explicit semantic cues to complement the implicit interaction signals.

\subsection{Task Definition: Sequence-Aware Explainable Recommendation}
\label{subsec:explanation_generation}
The first task is to generate a natural language explanation $E(u,i)$ for a user--item pair $(u,i)$ in a post-hoc manner, which justifies why item $i$ aligns with user $u$'s interests based on their sequential interaction history. Formally,
\begin{equation}
    E(u, i) \;=\; f\Bigl(S_{(u)}, \{d_j: j \in S_{(u)}\}, \,d_i\Bigr),
\end{equation}
where $f(\cdot)$ aims to encode both the temporal evolution of user preferences and the semantic relationships among items. 
Unlike prior works that rely solely on IDs or static embeddings~\cite{li2021personalized,li2023personalized,ma2024xreclargelanguagemodels}, we emphasize modeling the sequential nature of user behavior to generate contextually relevant and personalized explanations.

\subsection{Unified Evaluation Framework: Explanation Utility Assessment}
\label{subsec:unified_evaluation_framework}
To address the limitations of text-similarity-based evaluation, we introduce a unified framework that assesses explanation quality through their practical utility in recommendation scenarios. We define an explanation-enhanced recommender that leverages generated explanations:
\begin{equation}
    \hat{y}(u, i \mid E(u,i)) \;=\; g\Bigl(S_{(u)},\, E(u,i)\Bigr),
\end{equation}
where $g(\cdot)$ produces a relevance score for the pair $(u,i)$ conditioned on the explanation $E(u,i)$. This framework provides a task-based measure of explanation effectiveness: if $E(u,i)$ indeed captures the genuine rationale behind why $i$ suits $u$, then leveraging $E(u,i)$ as an additional semantic signal should improve prediction accuracy.

Importantly, this evaluation framework serves as a complement to traditional text metrics, not as a replacement for the primary explainable recommendation task. The framework enables us to assess whether explanations provide meaningful insights that could enhance user decision-making, offering a more comprehensive evaluation of explanation quality beyond linguistic similarity.

\begin{figure*}[t]
  \includegraphics[width=2.08\columnwidth]{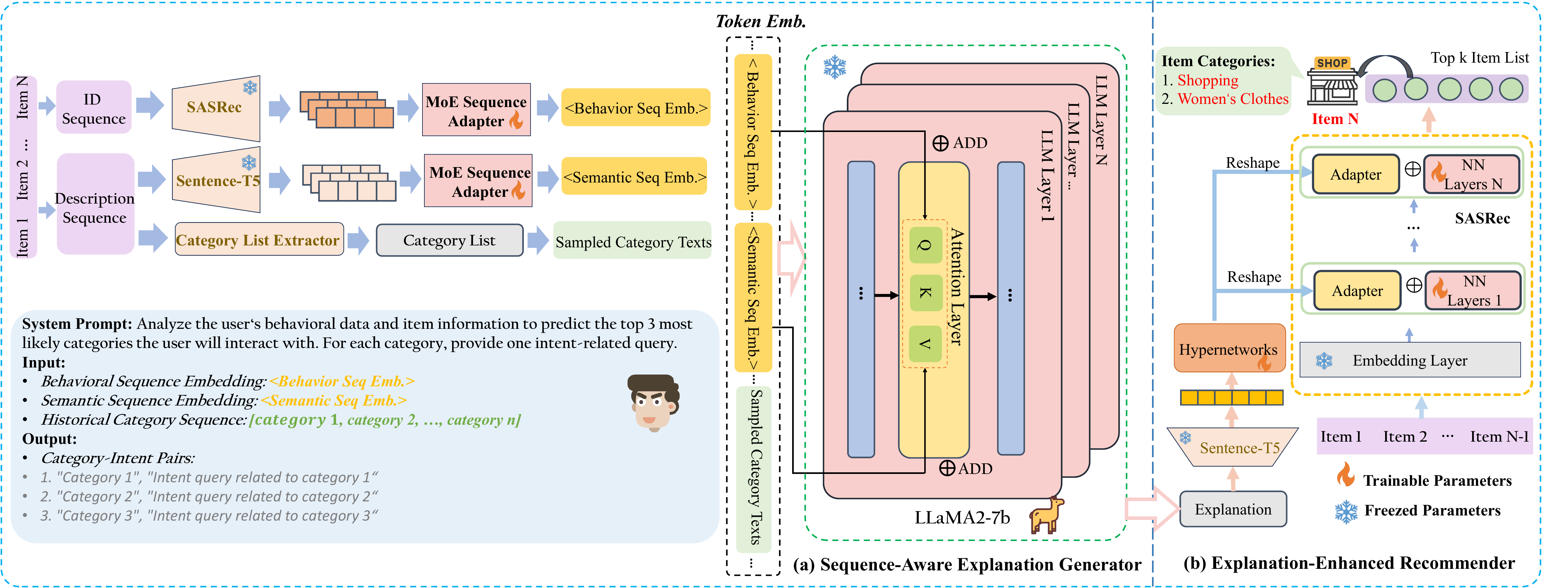}
  \caption{Overall architecture of the SELLER framework. The Sequence-Aware Explanation Generator (SEG) captures user behavioral patterns through dual-path encoding and generates personalized explanations via MoE-based adaptation. The Explanation-Enhanced Recommender (EER) serves as our unified evaluation framework to assess explanation utility through downstream recommendation performance.}
  \label{fig:framework}
\end{figure*}

\section{Methodology}
\label{sec:methodology}
Our proposed SELLER framework consists of two main components: (1) a \textbf{Sequence-Aware Explanation Generator (SEG)} that leverages user behavioral sequences and item semantic sequences to generate personalized explanations, and (2) a \textbf{unified evaluation framework} using an \textbf{Explanation-Enhanced Recommender (EER)} that assesses the practical utility of generated explanations. Figure \ref{fig:framework} illustrates the overall architecture of SELLER.

\medskip \noindent
\subsection{Sequence-Aware Explanation Generator}
\label{subsec:seg}
The SEG aims to generate high-quality, personalized explanations that capture the evolution of user preferences through their interaction history. 

\subsubsection{Dual-Path Sequence Encoding}
\label{subsubsec:dual_path}
Given a user $u$ and their interaction sequence $S_u = [i_1, i_2, \ldots, i_t]$, we propose a dual-path encoding scheme that captures complementary sequence information. Behavioral patterns reveal implicit user preferences through interaction histories, while semantic features extract explicit interest signals from item descriptions. This complementarity is crucial as behavioral patterns may miss semantic relationships between items, while pure semantic analysis may overlook temporal dynamics.

\medskip \noindent
\textbf{Behavioral Sequence Encoding.} We employ a pre-trained SASRec \cite{kang2018selfattentivesequentialrecommendation} model to encode the user's behavioral sequence:
\begin{equation}
    \mathbf{S}^{rec}_{u} = \phi_{rec}(S_u) \in \mathbb{R}^{d \times |S_u|}
\end{equation}
where $\phi_{rec}(\cdot)$ denotes the SASRec encoder and $d$ is the embedding dimensionality.

\medskip \noindent
\textbf{Semantic Sequence Encoding.} We utilize Sentence-T5 \cite{ni2021sentencet5scalablesentenceencoders} to encode the item descriptions corresponding to the user's interaction sequence:
\begin{equation}
    \mathbf{S}^{sem}_{u} = \phi_{sem}([d_{i_1}, d_{i_2}, \ldots, d_{i_t}]) \in \mathbb{R}^{d \times |S_u|}
\end{equation}
where $d_{i_j}$ represents the textual description of item $i_j$, and $\phi_{sem}(\cdot)$ is the Sentence-T5 encoder.

\subsubsection{MoE-based Sequence Adaptation}
\label{subsubsec:moe_adaptation}
To align the heterogeneous sequence representations with the LLM's semantic space, we introduce parallel MoE-based sequence adapters. For each representation path $p \in \{rec, sem\}$, the sequence embeddings are first processed by a Transformer encoder for temporal fusion:
\begin{equation}
    \mathbf{H}^{p} = \text{TransformerEncoder}(\text{LayerNorm}(\mathbf{S}^{p}_{u})).
\end{equation}
The sequence-level information is then obtained through mean pooling, followed by MoE-based adaptation:
\begin{equation}
    \mathbf{h}^{p} = \text{MeanPool}(\mathbf{H}^{p})
\end{equation}
\begin{equation}
    \mathbf{e}^{p}_{adapt} = \sum_{i=1}^g g_i(\mathbf{h}^{p}) \cdot \text{Expert}_i(\mathbf{h}^{p}),
\end{equation}
where $g_i(\cdot)$ represents the gating networks determining expert weights, and each expert is implemented as a two-layer feed-forward network. This process yields two adapted embeddings: $\mathbf{e}^{rec}_{adapt}$ for behavioral patterns and $\mathbf{e}^{sem}_{adapt}$ for semantic features.

\subsubsection{Sequence-Aware LLM Fine-tuning}
\label{subsubsec:llm_finetuning}
We enhance the LLM's generation capability by incorporating the adapted sequence representations into both input construction and attention computation. The input token sequence is structured as:
\begin{equation}
    \mathbf{X}^{LLM} = [\mathbf{x}^{LLM}_{sys}; \mathbf{e}^{rec}_{adapt}; \mathbf{e}^{sem}_{adapt}; \mathbf{x}^{LLM}_{cat}]
\end{equation}

where $\mathbf{x}^{LLM}_{sys}$ represents the fixed system prompt embedding, $\mathbf{e}^{rec}_{adapt}$ and $\mathbf{e}^{sem}_{adapt}$ are the MoE-adapted behavioral pattern and semantic feature embeddings, respectively, and $\mathbf{x}^{LLM}_{cat}$ denotes the embeddings of the sampled category texts from the item sequence. The inclusion of $\mathbf{x}^{LLM}_{cat}$ as an explicit text prompt provides richer context for the LLM without incurring significant computational overhead. The specific sampling method and examples are detailed in Appendix \ref{appendix:preprocessing}.


This design enables the model to dynamically adjust its attention patterns based on both the current textual context and the user's sequential behavior patterns. We inject sequence information directly into attention computation:

\begin{equation}
\begin{aligned}
    \mathbf{Q}_l &= \mathbf{Q}_{l}^{base} + \mathbf{e}^{rec}_{adapt} + \mathbf{e}^{sem}_{adapt} \\
    \mathbf{K}_l &= \mathbf{K}_{l}^{base} + \mathbf{e}^{rec}_{adapt} + \mathbf{e}^{sem}_{adapt} \\
    \mathbf{V}_l &= \mathbf{V}_{l}^{base} + \mathbf{e}^{rec}_{adapt} + \mathbf{e}^{sem}_{adapt}
\end{aligned}
\end{equation}

The attention output is computed as:
\begin{equation}
    \text{Attention}_l = \text{Softmax}\left(\frac{\mathbf{Q}_l\mathbf{K}_l^\top}{\sqrt{d}}\right)\mathbf{V}_l
\end{equation}

\paragraph{Training Objective} The SEG is optimized using a language modeling objective while keeping the backbone LLM parameters frozen. Given a user's complete interaction sequence $S_u = [i_1, i_2, \ldots, i_N]$ and the target item $i_N$, the model learns to generate explanations in a post-hoc manner:
\begin{equation}
\mathcal{L}_{exp} = -\sum_t \log P(y_t | y_{<t}, \mathbf{e}^{rec}_{adapt}, \mathbf{e}^{sem}_{adapt})
\end{equation}
where $y_t$ represents the $t$-th token in the ground truth explanation, and the model has access to the complete sequence including the target item $i_N$ during training.

\subsection{Explanation-Enhanced Recommender}
\label{subsec:evaluation_framework}
The EER serves as our unified evaluation framework to assess the practical utility of generated explanations through their impact on downstream recommendation performance.

\subsubsection{Explanation Encoding}
\label{subsubsec:explanation_encoding}

Given the generated explanation $E(u, i)$ for user $u$ and item $i$, we encode it into a dense representation using the Sentence-T5 encoder:
\begin{equation}
    \mathbf{e}_{explain} = \phi_{sem}(E(u, i))
\end{equation}

\subsubsection{Hypernetwork-Generated Adapter}
To make SASRec explanation-aware, we employ a hypernetwork-based approach rather than simple concatenation or addition of explanation embeddings. While direct feature concatenation can introduce explanation information, it may fail to capture complex interactions between explanations and user preferences. Our hypernetwork generates adapter parameters for SASRec's Transformer layers conditioned on the explanation embedding $\mathbf{e}_{explain}$, enabling more flexible and expressive explanation-preference integration through dynamic parameter adaptation.

For each Transformer layer $l$, the adaptive parameters are computed as:
\begin{equation}
    \mathbf{W}'_l = \mathbf{W}_l + f^{hyper}_l(\mathbf{e}_{explain})
\end{equation}

where $f^{hyper}_l$ is a layer-specific MLP. The generated parameters are reshaped to match the dimensions of the corresponding target parameters in SASRec's Transformer layers parameters and added to them. 

\subsubsection{Explanation-Aware SASRec Fine-tuning}
\label{subsubsec:eer_training}
During the independent training phase, the EER is fine-tuned using a modified BPR loss that incorporates both sequential information and explanation conditioning. Given a user sequence $S_u = [i_1, i_2, \ldots, i_{N-1}]$ and the target item $i_N$, the loss is computed as:
\begin{equation}
\begin{aligned}
    \mathcal{L}_{rec} = &-\sum \log \sigma\Big( \hat{y}(u, i_N | S_u, \mathbf{e}_{explain}^{GT}) \\
    &\quad - \hat{y}(u, i^- | S_u, \mathbf{e}_{explain}^{GT}) \Big)
\end{aligned}
\end{equation}
where $i^-$ is a randomly sampled negative item, $\mathbf{e}_{explain}^{GT}$ is the embedding of the ground truth explanation for item $i_N$, and $\hat{y}(u, i | S_u, e_{explain})$ represents the predicted relevance score that incorporates both the user's sequential behavior $S_u$ and the explanation condition $e_{explain}$.




\subsection{Training and Evaluation Protocol}
\label{subsec:training_evaluation}
The SEG and EER are trained independently to ensure fair evaluation. During training, SEG learns to generate post-hoc explanations given complete user sequences $S_u = [i_1, \ldots, i_N]$ including the target item, while EER learns next-item prediction using historical sequences $S_u = [i_1, \ldots, i_{N-1}]$ and ground truth explanation embeddings.

During evaluation, we employ a post-hoc protocol to assess explanation utility without information leakage: SEG generates explanations for test pairs $(u, i_N)$ using complete sequences, while EER predicts the target item $i_N$ using only the historical sequence $S_u = [i_1, \ldots, i_{N-1}]$ and the generated explanation embedding. This setup measures whether explanations capture meaningful user-item relationships by evaluating their predictive utility in an information-constrained setting. All baseline methods undergo the same evaluation protocol using our unified EER framework, ensuring fair comparison.

\subsection{Ground Truth Explanation Generator}
\label{subsec:ground_truth}

The ground truth explanations used for training are constructed using Claude 3.5 Sonnet, following common practices in explainable recommendation research. We generate two types of explanations:

\paragraph{Category-level Explanations} provide coarse-grained justifications directly linked to item metadata (e.g., the item belongs to ``Restaurants'').

\paragraph{Intent-level Explanations} use an LLM to generate plausible user intents within a category (e.g., for category ``Restaurants'', a potential intent query might be ``family-friendly dining options nearby'').

This dual-level approach captures both explicit categorical preferences and implicit user intentions, providing richer training signals for the explanation generation model. The specific construction methodology and examples are detailed in Appendix \ref{appendix:ground_truth}.


\section{Experiments}
We conduct extensive experiments to evaluate SELLER's effectiveness in both explanation generation quality and explanation utility assessment. 

\subsection{Experimental Settings}
\label{subsec:experimental_settings}

\paragraph{Datasets.} We evaluate the proposed SELLER framework on two widely-used datasets: Yelp\footnote{https://www.yelp.com/dataset/}  and KuaiRec\footnote{https://kuairec.com/} \cite{gao2022kuairec}. Yelp provides a rich collection of user interactions with businesses, while Kuairec offers a fully-observed user-item interaction matrix for short-form video recommendations. For both datasets, we apply 10-core filtering to retain interactions where both users and items have more than 10 interactions. Table~\ref{tab:data_statistics} summarizes the key statistics of the filtered datasets.

\begin{table}[h]
    \centering
    \caption{Dataset statistics.}
    \label{tab:data_statistics}
    \resizebox{\columnwidth}{!}{
    \begin{tabular}{lcccc}
        \toprule
        \textbf{Dataset} & \textbf{\#Users} & \textbf{\#Items} & \textbf{\#Interactions} & \textbf{\#Categories} \\
        \midrule
        Yelp    & 25,633  & 89,994  & 396,805 & 1253\\
        Kuairec & 6,584   & 8,180   & 342,700 & 446\\
        \bottomrule
    \end{tabular}
    }
\end{table}

\paragraph{Data Splitting.} We split the user interaction sequences chronologically into training, validation, and testing sets with a ratio of 8:1:1. The split is performed at a specific point in time to ensure a realistic evaluation setup and prevent future information leakage~\cite{sun2023Evaluating}. The maximum sequence length is set to 50 for all experiments.


\paragraph{Evaluation Metrics.}
We evaluate SELLER from two complementary perspectives:
\begin{enumerate}
\item \textbf{Explanation Generation Quality}: We employ BLEU~\cite{sellam2020bleurt} to measure lexical overlap and BERTScore~\cite{zhang2019bertscore} to assess semantic similarity between generated and reference explanations.
\item \textbf{Explanation Utility Assessment} We use our unified evaluation framework (Section~\ref{subsec:evaluation_framework}) to measure how effectively explanations help the EER predict target items. We report Recall@K and Normalized Discounted Cumulative Gain (NDCG@K), with K set to 5 and 10, to assess explanation utility through recommendation performance improvement.
\end{enumerate}

\paragraph{Baselines.} We compare SELLER against state-of-the-art explainable recommendation methods, including:
\begin{itemize}
\setlength{\itemsep}{0pt}     
\setlength{\topsep}{0pt}      
\setlength{\parsep}{0pt}      
\setlength{\partopsep}{0pt}
\item \textbf{PETER}~\cite{li2021personalized}: A post-hoc explanation generation approach that fine-tunes pre-trained language models using user and item IDs as input.
\item \textbf{PEPLER}~\cite{li2023personalized}: An explainable recommendation framework that incorporates user and item embeddings into pre-trained language models for explanation generation.
\item \textbf{XRec}~\cite{ma2024xreclargelanguagemodels}: A method that leverages both latent representations and explicit features to generate explanations using large language models.
\end{itemize}
Additionally, we include two control baselines:
\begin{itemize}
\item \textbf{Random}: Uses randomly sampled explanations from the training set.
\item \textbf{Empty}: Uses empty string explanations to isolate the effect of the adapter mechanism.
\end{itemize}

\begin{table*}[htbp]
  \centering
  \caption{Performance comparison of explanation utility assessment using the Explanation-Enhanced Recommender. The best results are highlighted in \textbf{bold}, and the second-best are \underline{underlined}. R@K and N@K denote Recall@K and NDCG@K, respectively.}
  
  \resizebox{1\textwidth}{!}{
    \begin{tabular}{l|rrrr|rrrr}
    \toprule
    \multirow{2}[2]{*}{\textbf{Method}} & \multicolumn{4}{c|}{\textbf{Yelp}} & \multicolumn{4}{c}{\textbf{Kuairec}} \\
          & \multicolumn{1}{c}{\textbf{Recall@5}} & \multicolumn{1}{c}{\textbf{Recall@10}} & \multicolumn{1}{c}{\textbf{NDCG@5}} & \multicolumn{1}{c|}{\textbf{NDCG@10}} & \multicolumn{1}{c}{\textbf{Recall@5}} & \multicolumn{1}{c}{\textbf{Recall@10}} & \multicolumn{1}{c}{\textbf{NDCG@5}} & \multicolumn{1}{c}{\textbf{NDCG@10}} \\
    \midrule
    \midrule
    \textbf{Random } & 0.0088  & 0.0156  & 0.0053  & 0.0075  & 0.1260  & 0.1960  & 0.0789  & 0.1015  \\
    \textbf{Empty} & 0.0072  & 0.0136  & 0.0046  & 0.0067  & 0.1272  & 0.1956  & 0.0795  & 0.1016  \\
    \textbf{PETER} & \underline{0.0116}  & 0.0144  & \underline{0.0067}  & 0.0076  & 0.1168  & 0.1832  & 0.0719  & 0.0935  \\
    \textbf{PEPLER} & 0.0100  & 0.0172  & 0.0062  & 0.0085  & 0.1252  & 0.1936  & 0.0814  & 0.1036  \\
    \textbf{XRec} & \underline{0.0116}  & \underline{0.0192}  & \underline{0.0067}  & \underline{0.0091}  
               & \underline{0.1280}  & \underline{0.1972}  & \underline{0.0817}  & \underline{0.1042}  \\

    \textbf{SELLER} & \textbf{0.0144} & \textbf{0.0228} & \textbf{0.0089} & \textbf{0.0116} & \textbf{0.1344} & \textbf{0.2132} & \textbf{0.0854} & \textbf{0.1110} \\
    \midrule
    \textbf{Improve} & 24.14\% & 18.75\% & 33.40\% & 27.52\% & 5.00\% & 8.11\% & 4.54\% & 6.57\% \\
    \bottomrule
    \end{tabular}%
    }
  \label{tab:explanation_utility}
\end{table*}%

\paragraph{Implementation Details.} We initialize the item embeddings with 64-dimensional vectors from SASRec, which consists of 2 Transformer layers. The Sentence-T5 encoder produces 768-dimensional representations. In the MoE-based sequence adaptation module, we employ a two-layer Transformer encoder with 64-dimensional embeddings, 4 attention heads. The MoE module comprises 8 experts , each implemented as a two-layer feedforward network with a dropout rate of 0.2. We adopt LLaMA2-7B \cite{Touvron2023Llama2O} as the language language model for explanation generation. The Sequence-Aware Explanation Generator (SEG) is trained for 5 epochs using the Adam optimizer with a batch size of 8, a learning rate of 1e-4, and a weight decay of 1e-6. The Explanation-Enhanced Recommender (EER) is trained for 50 epochs using the Adam optimizer with a batch size of 256, a learning rate of 5e-4, and a weight decay of 1e-5.

\subsection{Results and Analysis}

\paragraph{Explanation Utility Assessment.}

Table~\ref{tab:explanation_utility} presents the utility assessment of explanations generated by different ER methods within our unified evaluation framework. 

\begin{itemize}
\item SELLER achieves superior explanation utility across both datasets. Compared with XRec, SELLER obtains 18.75\% and 8.11\% improvements in Recall@10, and 27.52\% and 6.57\% improvements in NDCG@10 on Yelp and KuaiRec, respectively. This demonstrates that SELLER's sequence-aware explanations provide more useful semantic signals for the recommendation task.

\item The significant relative improvements observed between SELLER and baseline methods within this consistent evaluation framework demonstrate the superior quality and practical relevance of explanations generated by our sequence-aware approach. While absolute metrics reflect the specific evaluation setup, the comparative gains robustly indicate explanation quality differences.

\item Baseline performance variations across datasets reveal important insights. On Yelp, all ER methods outperform ``Random'' and ``Empty'' baselines, validating the utility of personalized explanations. However, on KuaiRec, PEPLER and PETER underperform, suggesting that dense interaction sequences require more sophisticated explanation generation models.

\end{itemize}


\begin{table}[htbp]
  \centering
  \caption{Explanation quality comparison.} 
  \resizebox{0.9\columnwidth}{!}{
    \begin{tabular}{l|cc|cc}
    \toprule
    \multirow{2}[2]{*}{\textbf{Method}} & \multicolumn{2}{c|}{\textbf{Yelp}} & \multicolumn{2}{c}{\textbf{Kuairec}} \\
          & \textbf{BLEU} & \textbf{BERTScore} & \textbf{BLEU} & \textbf{BERTScore} \\
    \midrule
    \midrule
    \textbf{PETER} & 0.1242  & 0.6088  & 0.1121  & 0.5184  \\
    \textbf{PEPLER } & 0.2632  & 0.7063  & 0.1123  & 0.5770  \\
    \textbf{XRec} & \underline{0.3243}  & \underline{0.7328}  & \underline{0.1536}  & \underline{0.6265}  \\
    \textbf{SELLER} & \textbf{0.3815}  & \textbf{0.7524}  & \textbf{0.2072}  & \textbf{0.7011}  \\
    \midrule
    \textbf{Improve} & 17.64\% & 2.67\% & 34.90\% & 11.91\% \\
    \bottomrule
    \end{tabular}%
    }
  \label{tab:explanation_quality}
\end{table}%

\paragraph{Explanation Generation Quality.} Table~\ref{tab:explanation_quality} 
evaluates explanation quality using traditional text similarity metrics, providing an intrinsic assessment complementary to the utility-based evaluation.

\begin{itemize}
\setlength{\itemsep}{0pt}     
\setlength{\topsep}{0pt}      
\setlength{\parsep}{0pt}      
\setlength{\partopsep}{0pt}

\item  SELLER generates explanations with the highest lexical and semantic similarity to the groundtruth on both datasets. SELLER obtains larger improvements on the Kuairec dataset, which contains dense interaction sequences. This verifies the superiority of SELLER in capturing sequential information for explanation generation.
\item Interestingly, while SELLER obtains larger gains on Kuairec in terms of text similarity, its improvements in recommendation performance are more significant on Yelp. This implies that higher textual similarity to the groundtruth explanations does not necessarily indicate better reflection of users' real decision-making process. The groundtruth explanations themselves may contain noise and bias. Therefore, evaluating explanation quality solely based on text similarity can be unreliable, further emphasizing the importance of the integrated evaluation approach adopted by SELLER.
\end{itemize}

\paragraph{Combined Analysis.} The complementary nature of Tables~\ref{tab:explanation_utility} and~\ref{tab:explanation_quality} provides a comprehensive view of explanation quality. While Table~\ref{tab:explanation_quality} measures linguistic fidelity, Table~\ref{tab:explanation_utility} assesses practical utility, together offering a more complete evaluation than either metric alone. SELLER's consistent superiority across both evaluations validates its effectiveness in sequence-aware explainable recommendation.

\subsection{Ablation Study}

\begin{figure}[h]
\centering
\includegraphics[width=0.48\textwidth]{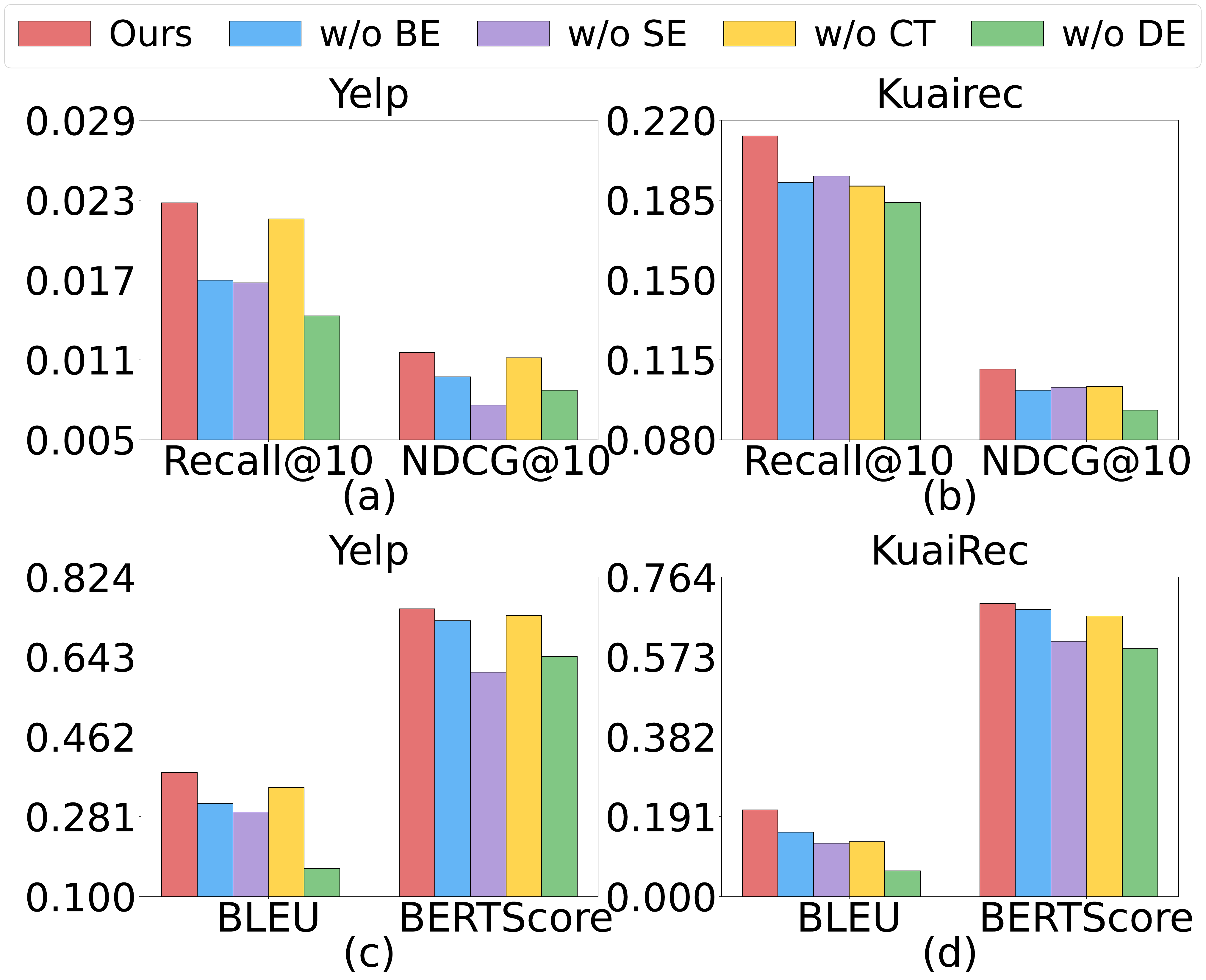}
\caption{Ablation study on the sequence-aware components of SELLER. BE: MoE-adapted behavioral embedding, SE: MoE-adapted semantic embedding, CT: category texts, DE: BE+SE.}
\label{fig:ablation}
\end{figure}

\paragraph{Impact of Sequence-Aware Components.}
To validate the effectiveness of our sequence-aware design, we conduct ablation studies by removing key components:

\begin{itemize}
\setlength{\itemsep}{0pt}     
\setlength{\topsep}{0pt}      
\setlength{\parsep}{0pt}      
\setlength{\partopsep}{0pt}
\item \textbf{w/o BE}: SELLER without MoE-adapted behavioral embedding ($\mathbf{e}^{rec}_{adapt}$)
\item \textbf{w/o SE}: SELLER without MoE-adapted semantic embedding ($\mathbf{e}^{sem}_{adapt}$)
\item \textbf{w/o CT}: SELLER without category texts ($\mathbf{x}^{LLM}_{cat}$)
\item \textbf{w/o DE}: SELLER without both behavioral and semantic embeddings
\end{itemize}

As shown in Figure~\ref{fig:ablation}, removing any component leads to performance degradation, with the following key observations:

\begin{itemize}
\setlength{\itemsep}{0pt}     
\setlength{\topsep}{0pt}      
\setlength{\parsep}{0pt}      
\setlength{\partopsep}{0pt}
\item The significant drop in w/o DE validates our motivation to capture both implicit behavioral patterns and explicit semantic features for explanation generation.
\item On Yelp, w/o SE causes larger performance drops than w/o BE , suggesting that semantic features from detailed reviews provide stronger signals for explanation generation in the Yelp domain.
\item On Kuairec, w/o BE and w/o SE show comparable impact, indicating that both behavioral and semantic sequences contribute similarly to short video recommendations where user interactions are more frequent and diverse.
\end{itemize}

\begin{table}[htbp]
  \centering
  \caption{Performance comparison with different explanation inputs for the recommendation model.}
  \setlength{\tabcolsep}{2pt}  
  \renewcommand{\arraystretch}{1.3}  
  \resizebox{0.48\textwidth}{!}{
    \begin{tabular}{l|cc|cc}
    \toprule
    \multirow{2}[2]{*}{\textbf{Method}} & \multicolumn{2}{c|}{\textbf{Yelp}} & \multicolumn{2}{c}{\textbf{Kuairec}} \\
          & \textbf{Recall@10} & \textbf{NDCG@10} & \textbf{Recall@10} & \textbf{NDCG@10} \\
    \midrule
    \midrule
    \textbf{Random Category} & 0.0156  & 0.0075  & 0.1760  & \underline{0.1015}  \\
    \textbf{Category-level} & 0.0188  & 0.0096  & 0.1748  & 0.0920  \\
    \textbf{Intent-level} & \underline{0.0192}  & 0.0093  & \underline{0.1892}  & 0.0984  \\
    \textbf{Both (SELLER)} & \textbf{0.0228}  & \textbf{0.0116}  & \textbf{0.2132}  & \textbf{0.1110}  \\
    \bottomrule
    \end{tabular}%
  }
  \label{tab:explanation_types}
\end{table}%

\paragraph{Effect of Explanation Types.} To evaluate our hypernetwork-based explanation injection mechanism, we compare different types of explanatory inputs for recommendation (Table~\ref{tab:explanation_types}):

\begin{itemize}
\item All explanation variants outperform the random text baseline, validating the effectiveness of our hypernetwork-based adapter in utilizing explanatory signals for recommendation.
\item Intent-level explanations achieve better performance than category-level ones, suggesting that fine-grained semantic information provides stronger guidance for recommendation.
\item  The superior performance of combining both explanation types demonstrates our model's capability to effectively integrate complementary semantic signals through the hypernetwork mechanism.
\end{itemize}

\paragraph{Efficiency Analysis.} We compare the training and inference efficiency of SELLER with baselines. As shown in Table~\ref{tab:efficiency}, SELLER has a much lower training and inference cost compared with XRec, while slightly higher than LLaMA2. This is because SELLER does not incorporate additional user or item profile texts, leading to shorter input and lower computational cost. The results demonstrate the efficiency of SELLER in both training and inference.


\begin{table}[h]
    \centering
    \resizebox{0.4\textwidth}{!}{
        \begin{tabular}{lcc}
        \toprule
        \textbf{Method} & \textbf{Training Time} & \textbf{Inference Time} \\
        \midrule  
        XRec & 1$\times$ & 1$\times$ \\
        SELLER & \textbf{0.735}$\times$ & 0.448$\times$ \\
        LLaMA2-7B & \textit{---} & \textbf{0.328}$\times$ \\
        \bottomrule
        \end{tabular}
    }
    \caption{Comparison of training and inference efficiency. The training time and inference time are relative to XRec.}
    \label{tab:efficiency}
\end{table}



\section{Related Work}
In this section, we review relevant literature on explainable recommendation systems and parameter-efficient fine-tuning, highlighting the gaps our work addresses.

\subsection{Explainable Recommendation Systems}
Early approaches to explainable recommendation focused on extracting explanations from user reviews \cite{zhang2014explicit} or mining patterns from user-item interactions \cite{liu2023Triple} and knowledge graphs \cite{zhu-etal-2021-faithfully}. Template-based methods \cite{Zhang2018CRS, lei2020interactive} generated explanations by filling predefined templates with relevant features, while attention-based approaches \cite{Cong2019ICMR} leveraged attention mechanisms to highlight important factors influencing recommendations. However, these methods struggle with scalability and require extensive domain knowledge.

The emergence of large language models has opened new possibilities for generating more natural and contextually relevant explanations. PETER \cite{li2021personalized} pioneered the use of pre-trained language models for explanation generation by incorporating user and item IDs into the input prompt. PEPLER \cite{li2023personalized} extended this approach by injecting learned user and item embeddings into the LLM architecture. XRec \cite{ma2024xreclargelanguagemodels} further enhanced explanation quality by combining both latent collaborative filtering signals and explicit features.
While these LLM-based methods have significantly improved explanation quality, they primarily focus on static representations and treat explanation generation as a standalone task. Our work differs by explicitly modeling sequential patterns and proposing a unified framework that evaluates explanations based on their impact on recommendation performance.

\subsection{Parameter-Efficient Fine-tuning and Adaptation}
Parameter-efficient fine-tuning has emerged as a crucial technique for adapting large pre-trained models. Adapter modules \cite{pmlr-v97-houlsby19a} and LoRA \cite{Hu2021LoRALA} demonstrate the effectiveness of introducing small trainable components while keeping base models frozen. In recommendation systems, these methods have been applied to inject collaborative filtering signals into LLMs \cite{ma2024xreclargelanguagemodels}. However, existing approaches typically focus on aligning static embeddings, neglecting the sequential nature of user behaviors. The Mixture-of-Experts (MoE) architecture \cite{mustafa2022multimodalcontrastivelearninglimoe} has shown success in conditional computation and dynamic feature fusion. Hypernetworks \cite{lv-etal-2024-hyperlora} offer flexible parameter generation conditioned on input features, enabling efficient cross-task adaptation. While these techniques have been explored for LLM adaptation, their potential for sequence-aware explainable recommendation remains unexplored.

\section{Conclusion}
In this work, we propose SELLER, a novel framework for sequence-aware explainable recommendation that addresses two fundamental limitations of existing approaches: the neglect of temporal user behavior patterns and the lack of robust evaluation methodologies. Our contributions are twofold. First, we introduce a Sequence-Aware Explanation Generator that captures dynamic user preferences through dual-path encoding and MoE-based adaptation, enabling personalized explanations that reflect sequential interaction patterns. Second, we propose a unified evaluation framework that assesses explanation quality through their practical utility in recommendation tasks, complementing traditional text similarity metrics. Experiments on two public datasets demonstrate SELLER's effectiveness, achieving significant improvements in both explanation quality over state-of-the-art methods, while maintaining superior computational efficiency. These results highlight the importance of capturing sequential patterns in explainable recommendation and offer a principled approach to assess explanation utility beyond textual metrics, ultimately moving toward more transparent recommender systems that provide users with meaningful insights into their decision-making processes.

\section*{Limitations}
\label{sec:limitations}
Despite SELLER's advances, several limitations warrant future investigation.
\paragraph{Ground Truth Construction} Our approach relies on LLM-generated ground truth explanations, which may not fully capture users' genuine sequential intents. This item-side inference introduces potential biases compared to directly elicited user rationales. Future work should explore methods for obtaining more authentic ground truth explanations or developing approaches less dependent on labeled references.
\paragraph{Evaluation Constraints} The current evaluation relies exclusively on offline metrics, which may not fully align with real-world user perceptions. Incorporating online user studies would provide deeper insights into how sequence-aware explanations impact user satisfaction and trust in practical scenarios.
\paragraph{Modality and Domain Coverage} SELLER primarily integrates textual and sequential behavioral signals, leaving multimodal explanations unexplored. Additionally, while evaluated on business and video recommendations, its effectiveness in other domains requires further validation. Future research should explore cross-modal and cross-domain adaptations to enhance the framework's generalizability.





\bibliography{latex/acl_latex}

\appendix
\newpage

\section{Appendix}
\label{sec:appendix}

In this appendix, we provide additional details on: (A) data preprocessing for the SELLER framework, (B) construction of ground truth explanations, and (C) a comprehensive case study demonstrating the effectiveness of our sequence-aware explainable recommendation approach.

\subsection{Data Preprocessing for Sequence-Aware LLM Input}
\label{appendix:preprocessing}
The effectiveness of our Dual-Path Sequence Encoding module relies on properly structured input data for the Language Model (LLM). We detail the preprocessing steps that transform raw interaction data into semantically rich representations.

\subsubsection{Item Description Construction}
\label{appendix:item_desc}

\begin{figure}[!t]
\centering
\includegraphics[width=0.97\columnwidth]{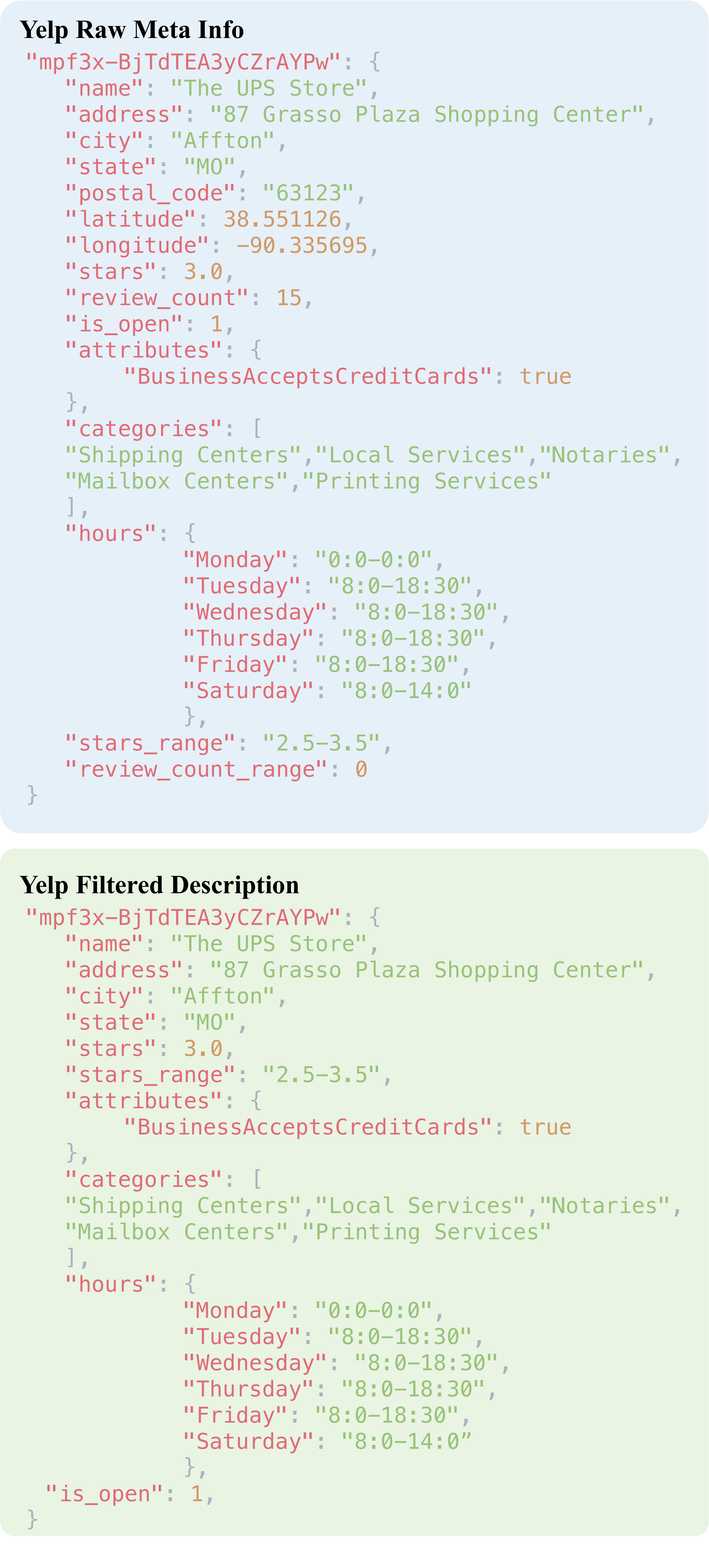}
\caption{Structured meta-information filtering process for the Yelp dataset. The raw metadata is filtered to retain relevant fields that capture business characteristics while removing non-essential information.}
\label{fig:yelp_filtering}
\end{figure}

For each item in a user's interaction sequence, we construct structured textual descriptions as follows:

\paragraph{Yelp Dataset.} We extract and combine key metadata fields including: name, address, city, state, star rating, star range, business attributes, categories, operating hours, and open status. Figure ~\ref{fig:yelp_filtering} illustrates this filtering process, where non-essential fields (e.g., postal code, latitude, longitude) are removed, while preserving information that captures geographical, categorical, and temporal relevance.

\paragraph{KuaiRec Dataset.} Due to limited metadata in this short-form video recommendation dataset, we primarily use the category field as the item's textual description to maintain consistency across datasets.

\subsubsection{Category Sequence Construction}
\label{appendix:cat_sequence}

To capture high-level semantic patterns in user behavior, we construct category sequences as follows:

\begin{enumerate}
    \item Extract up to three categories per item from the item's category list
    \item Apply a word2vec-inspired downsampling strategy \cite{mikolov2013efficientestimationwordrepresentations} to balance category distribution while preserving diversity
\end{enumerate}

The downsampling probability is calculated as:
\[
P_{	ext{sampled}} = \begin{cases} 
1 - \left(\frac{t}{\text{freq}}\right)^{0.5}, & \text{if } \text{freq} > t \\ 
1.0, & \text{otherwise}
\end{cases}
\]
where $t$ is a threshold parameter (set to $1 \times 10^{-5}$ in our experiments) and $freq$ is the normalized frequency of the category in the dataset. This approach helps balance frequently occurring categories (e.g., "Restaurants" in Yelp) with more specific ones, leading to more diverse and informative category sequences.




\subsubsection{LLM Input Structure}
\label{appendix:llm_input}

\begin{figure}[!t]
  \centering
    \includegraphics[width=1\columnwidth]{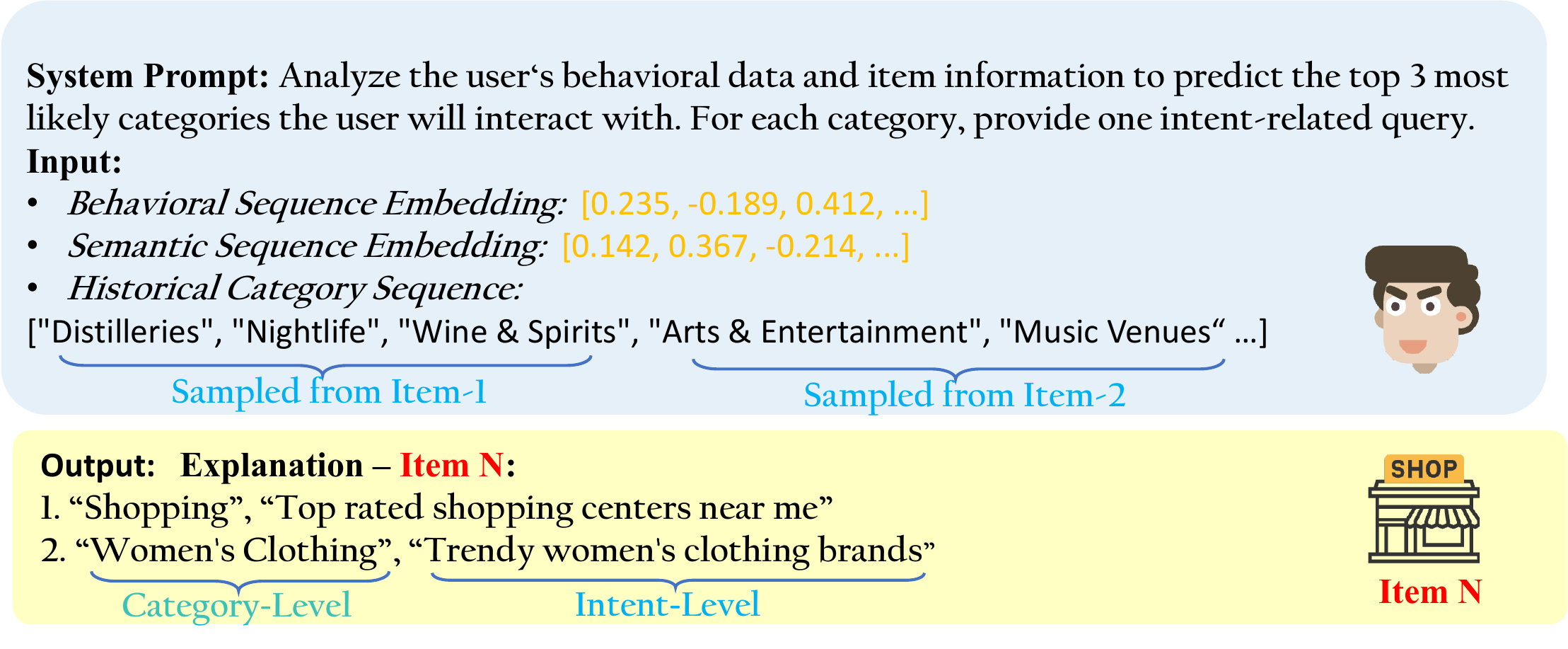}
  \caption{Input structure for the Sequence-Aware Explanation Generator, including the system prompt, input fields, and expected output format with examples.}
  \label{fig:llm_input}
\end{figure}

For the Sequence-Aware Explanation Generator, we format the input prompt structure as shown in Figure~\ref{fig:llm_input}. This system prompt guides the LLM to analyze user behavioral patterns based on the sequence representations and generate category-intent pairs that capture both explicit categories and underlying user intentions.

\subsection{Ground Truth Explanation Construction}
\label{appendix:ground_truth}

To train our Sequence-Aware Explanation Generator, we need high-quality ground truth explanations. Following recent work in LLM-based recommendations \cite{ma2024xreclargelanguagemodels}, we leverage Claude 3.5 Sonnet to generate two types of ground truth explanations:

\begin{enumerate}
    \item \textbf{Category-level explanations}: Coarse-grained justifications directly linked to item metadata
    \item \textbf{Intent-level explanations}: Fine-grained rationales capturing potential user intentions
\end{enumerate}

\subsubsection{Prompt Design for Ground Truth Generation}
\label{appendix:gt_prompts}

\begin{figure}[!t]
  \centering
    \includegraphics[width=1\columnwidth]{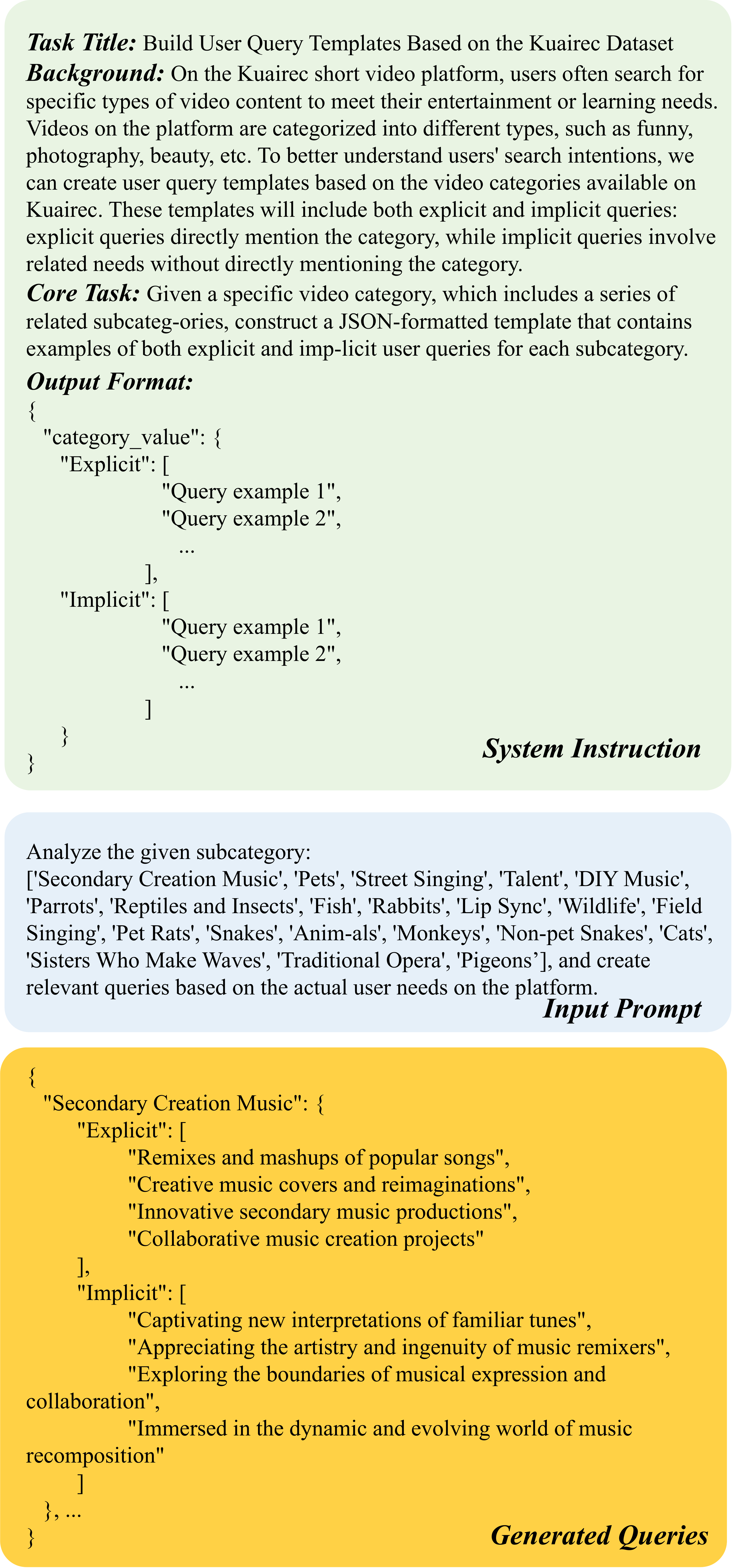}
  \caption{Ground truth explanation generation process for the KuaiRec dataset, showing the task specification, prompt design, and sample output format.}
  \label{fig:groundtruth_explanations}
\end{figure}

Figure~\ref{fig:groundtruth_explanations} illustrates our approach for generating ground truth explanations using the KuaiRec dataset. We carefully design the LLM prompt to elicit both explicit category-based explanations and implicit intent-based explanations.

In this design:

\begin{itemize}
    \item \textbf{Explicit queries} directly mention the category (e.g., "Remixes and mashups of popular songs" for "Secondary Creation Music")
    \item \textbf{Implicit queries} describe user needs without explicitly naming the category (e.g., "Captivating new interpretations of familiar tunes")
\end{itemize}

This dual-query approach enhances the model's ability to infer intent beyond direct keyword matching, accommodating both straightforward and nuanced user preferences. The generated explanations serve as valuable training data for the SELLER framework, enabling it to produce high-quality, contextually relevant explanations for recommended items.

\subsection{Case Study}
\label{appendix:case_study}

\begin{figure*}[t]
  \centering
    \includegraphics[width=2\columnwidth]{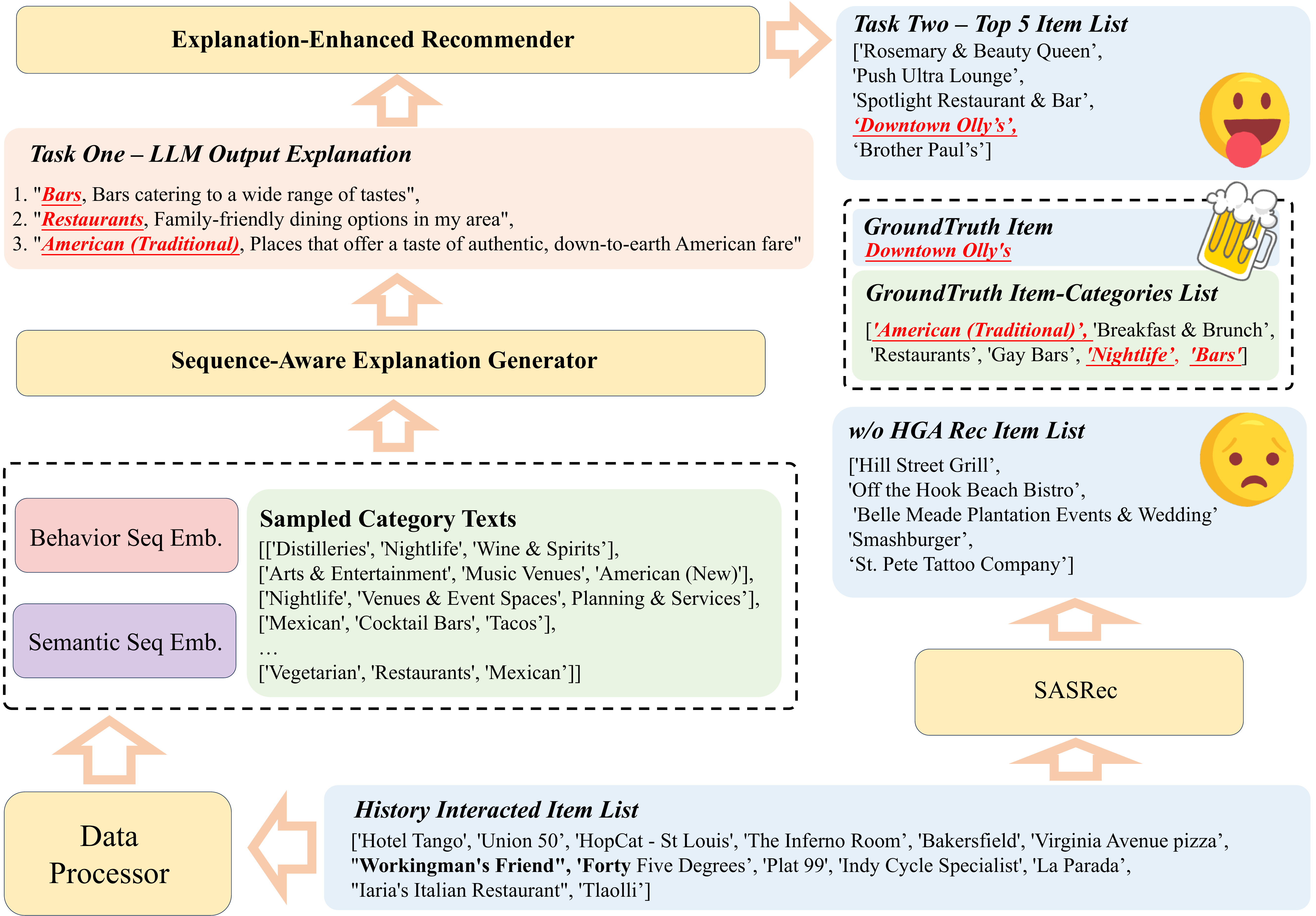}
  \caption{A comprehensive case study demonstrating SELLER's explanation generation capabilities. The framework leverages sequence information to generate contextually relevant explanations that capture the user's evolving preferences. The evaluation through our EER framework shows that these explanations successfully reflect meaningful user-item relationships.}
  \label{fig:case_study}
\end{figure*}

To illustrate the effectiveness of the SELLER framework for sequence-aware explainable recommendation, we present a detailed case study in Figure~\ref{fig:case_study}. This example demonstrates how our approach generates high-quality explanations that capture temporal patterns in user preferences through the Sequence-Aware Explanation Generator (SEG), and how these explanations are evaluated through our proposed Explanation-Enhanced Recommender (EER).

In this example:

\begin{enumerate}
    \item The user has interacted with various bars and restaurants in the past, as shown in the "History Interacted Item List"
    \item The Sequence-Aware Explanation Generator processes both behavioral and semantic sequence embeddings, along with sampled category texts
    \item The generator produces three key explanations that reflect the user's evolving preferences: "Bars catering to a wide range of tastes", "Family-friendly dining options in my area", and "Places that offer authentic, down-to-earth American fare"
    \item These explanations successfully capture the categories present in the target item "Downtown Olly's" (American Traditional, Restaurants, Bars)
    \item Our Explanation-Enhanced Recommender, as part of the evaluation framework, confirms the utility of these explanations by successfully ranking the target item in the top recommendations
\end{enumerate}

This case study highlights two key aspects of our framework. First, it demonstrates that SELLER can generate explanations that effectively capture the sequential evolution of user preferences, something that standard static approaches would miss. The explanations reflect not just general user preferences but the specific trajectory of interests over time.

Second, it validates our unified evaluation approach. The fact that our explanation-enhanced recommender can successfully identify the target item suggests that the generated explanations contain meaningful information about the user-item relationship. This supports our argument that explanation quality should be evaluated not just through text similarity metrics but through their practical utility in capturing genuine recommendation rationales.

\end{document}